\RequirePackage{amsmath}
\documentclass[runningheads]{llncs}
\usepackage[T1]{fontenc}
\usepackage{graphicx}
\usepackage{bbm, mathtools}
\usepackage{color}
\usepackage{tabularx}
\usepackage{soul}
\usepackage{mwe}
\usepackage{float}
\usepackage{enumitem} 
\usepackage{makecell} 
\usepackage{breqn}
\PassOptionsToPackage{hyphens}{url}\usepackage{hyperref}
\usepackage{marvosym}

\begin{document}
\raggedbottom
\title{Nonrigid Reconstruction of Freehand Ultrasound without a Tracker}
\titlerunning{Nonrigid trackerless freehand ultrasound}
%

%
\author{
Qi Li\inst{1}\textsuperscript{(\Letter)}\and
%
Ziyi Shen\inst{1} \and
Qianye Yang\inst{1} \and
Dean C. Barratt\inst{1} \and
Matthew J. Clarkson\inst{1} \and
Tom Vercauteren\inst{2} \and
Yipeng Hu\inst{1}}

\authorrunning{Q. Li et al.}
\institute{Centre for Medical Image Computing, Wellcome/EPSRC Centre for Interventional and Surgical Sciences, Department of Medical Physics and Biomedical Engineering, University College London, London, U.K.\\
\email{qi.li.21@ucl.ac.uk}\and
School of Biomedical Engineering \& Imaging Sciences, King’s College London, London, U.K.
}

\maketitle              
\begin{abstract}
Reconstructing 2D freehand Ultrasound (US) frames into 3D space without using a tracker has recently seen advances with deep learning.
Predicting good frame-to-frame rigid transformations is often accepted as the learning objective, especially when the ground-truth labels from spatial tracking devices are inherently rigid 
transformations. Motivated by a) the observed nonrigid deformation due to soft tissue motion during scanning, and b) the highly sensitive prediction of rigid transformation, 
this study investigates the methods and their benefits in predicting nonrigid transformations for reconstructing 3D US. 
We propose a novel co-optimisation algorithm for simultaneously estimating rigid transformations among US frames, supervised by ground-truth from a tracker, and a nonrigid deformation, optimised by a regularised registration network. We show that these two objectives can be either optimised using meta-learning or combined by weighting. A fast scattered data interpolation is also developed for enabling frequent reconstruction and registration of non-parallel US frames, during training.   
With a new data set containing over 357,000 frames in 720 scans, acquired from 60 subjects, the experiments demonstrate that, due to an expanded thus easier-to-optimise solution space, the generalisation is improved with the added deformation estimation, with respect to the rigid ground-truth. The global pixel reconstruction error (assessing accumulative prediction) is lowered from 18.48 to 16.51 mm, compared with baseline rigid-transformation-predicting methods. Using manually identified landmarks, the proposed co-optimisation also shows potentials in compensating nonrigid tissue motion at inference, which is not measurable by tracker-provided ground-truth. 
The code and data used in this paper are made publicly available at \url{https://github.com/QiLi111/NR-Rec-FUS}.

\end{abstract}

\section{Introduction}
\label{sec:intro}
With a variety of clinical applications including measurement assessment \cite{leblanc2022stretched}, pre-operative registration \cite{lang2012multi} and surgical guidance \cite{lindseth2003multimodal}, trackless freehand US reconstruction has been proposed using both non-learning \cite{chen1997determination} and machine learning-based approaches~\cite{prevost20183d}.
Recent learning-based methods vary in their network architectures \cite{mikaeili2022trajectory,xie2021image,ning2022spatial}, training strategies \cite{luo2021self,guo2022ultrasound,luo2023recon}, the use of prior knowledge \cite{luo2022deep,luo2023multi}, input frames
\cite{miura2020localizing,xie2021image} and sequential modelling techniques \cite{li2023trackerless,miura2021probe}.

To our knowledge, most existing approaches optimise rigid transformations among US frames,
which characterises the spatial movement - rotation and translation of the ultrasound probe during scanning. However, there is evidence that probe pressure and patient movement cause soft-tissue undergoing nonrigid deformation \cite{wein2020three}.
Previous work compensated this nonrigid deformation using a separate registration algorithm, after a rigid reconstruction \cite{prevost20183d}, aligning the same anatomical structures appeared in repeated scans \cite{wein2020three}.

However, validating such a nonrigid deformation modelling is challenging due to a lack of general means to obtain soft-tissue-tracking, motion-included ground-truth. In fact, the ground-truth data for supervision used in many learning-based approaches are obtained from spatial tracking devices, such as optical and electromagnetic trackers. They are rigid transformations that localise the rigid probe rather than the deformable anatomical structures. 

Withstanding the challenges in validating the estimated nonrigid deformation, we would like to explore other benefits for predicting nonrigid transformation in reconstructing US images.
Two types of other applications that require predicting rigid transformation are image registration and image data augmentation/perturbation in training neural networks. Both have reported the difficulties in esimating this constrained transformation. For example, predicting rigid transformation is highly sensitive to initialisation and learning rate in weakly-supervised registration \cite{hu2018weakly} and spatial transformer network training \cite{NIPS2015_33ceb07b}, mandating careful hyperparameter tuning in these applications. Predicting a higher degree-of-freedom, flexible nonrigid transformation provides an expanded solution space in optimising these transformation estimation methods, with respect to either rigid or nonrigid ground-truth labels. This should improve the resulting model generalisation, given limited data and compute resources in practice. 

In this paper, we propose a co-optimisation deep-learning-based approach, together with a ``conventional'' learnable rigid reconstruction, to estimate an additional nonrigid deformation between US frames as well as within acquired individual frames. Although the latter may not be plausibly compensate physical deformation due to the fast single-frame imaging process, this should allow the flexibility that benefits the numerical training process discussed above.  

In summary, our contributions include: 
1) a novel co-optimisation approach, not only for compensating nonrigid soft tissue motion but also improve the network training for better generalisation based on rigid ground-truth;
2) an open-sourced PyTorch implementation of a practical interpolation method for scatter ultrasound intensity values;
3) a new set of evaluation metrics for reconstruction evaluation, at both global and local levels;
and 4) one of the largest \textit{in vivo} US dataset for freehand US reconstruction, with recorded tracker information.

\section{Method}
\label{sec:method}
\begin{figure}[t]
 \centering
\includegraphics[width=\textwidth]{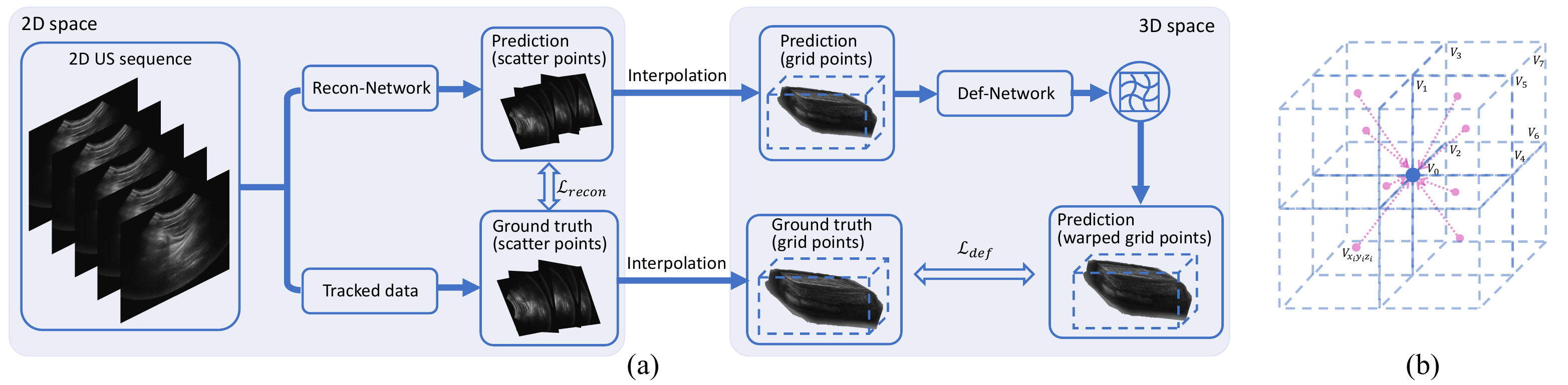}
\caption{(a) Overview of the proposed method with rigid transformation and nonrigid deformation prediction. (b) Contributions from support data to query data.
} \label{pipeline}
\end{figure}

An US sequence $S$ consists of a number of US frames $S=\{I_m\}, m=1,2,...,M$, acquired with increasing timestamps. For any pair of US frames $I_i$ and $I_j$, a spatial rigid transformation parameter vector $\textbf{t}^{j\leftarrow i}, 1 \leq i<j \leq M$ denotes the relative translation and rotation between the $i^{th}$ and $j^{th}$ frames.
An US scan containing several US sequences can be reconstructed in 3D once all the transformations between each US frame and the reference frame can be calculated, where the reference frame can be any frame in the scan. Fig.~\ref{pipeline} (a) provides an overview of the proposed method. 

\subsection{Rigid Reconstruction of Stacked 2D Frames}
\label{ssec:Reconstruction of stacked 2D frames}

\sloppy This section describes a rigid transformation estimation pipeline as proposed in \cite{prevost20183d}, using a deep neural network. 
As described in the sequence-modelling methods \cite{li2023trackerless}, a convolutional neural network $f_{recon}$ with parameters $\theta_{recon}$ predicts rigid transformations with respect to the reference frame:
$ [(\hat{\textbf{t}}_{1}^{ref\leftarrow 1})^\top, ..., (\hat{\textbf{t}}_{M}^{ref\leftarrow M})^\top] = f_{recon}(S;\theta_{recon}) $, 
where $\{\hat{\textbf{t}}_m^{ref\leftarrow m}\}_{m=1}^M$, is a set of rigid transformation parameter vectors from all frames to the reference frame.
The 3D coordinates of $n^{th}$ pixel in $m^{th}$ frame $\hat{P^n_m}$, in reference frame coordinate system, can be calculated using $\hat{P^n_m}=\mathcal{T}(P^n_{I_{0}},\hat{\textbf{t}}_{m}^{ref\leftarrow m})$ .
$P^n_{I_{0}}$ is the point coordinates in it's tracker tool coordinate system, and can be obtained by using the spatial calibration matrix \cite{hu2017freehand} $\textbf{t}_{calib}$: $P^n_{I_{0}}=\mathcal{T}(P^n_{img},{\textbf{t}}_{calib})$, where $P^n_{img}$ denotes the coordinates in the pixel coordinate system \cite{xingfang2010simple}. $\mathcal{T}$ is a function applies the $\hat{\textbf{t}}_{m}^{ref\leftarrow m}$-converted transformation on the points. 
While any frame in an US sequence can be regarded as the reference frame, we use the first frame as the reference frame.

\subsection{Interpolation for Efficient Resampling Non-parallel Frames}
\label{ssec:Interpolation Approximation}

In this section, we first describe an efficient approach for interpolating from scatter data to grid samples.
US intensities sampled at regular grid are useful to efficiently compute similarity measures, deformation regularisation and image warping, as commonly adopted in training registration networks. These are repeatedly applied in the proposed co-optimisation, described in Sec.~\ref{ssec:Co-Optimisation in 3D Space}. Many interpolation methods for estimating grid data from scatter data \cite{huang2012grid,amidror2002scattered,lee1997scattered,bartier1996multivariate,vercauteren2006robust,ebner2017point} are computationally expensive. We introduce a simple and efficient interpolation, with a time complexity of $O(N)$ where $N$ is number of pixels in a scan.
 
Denote intensity value at any point $(x,y,z)$ as $V_{x,y,z}$, where the coordinate system is defined such that the reconstructed voxel grids are positioned at integer coordinates. Suppose there are $N$ support data points at $\{(x_i, y_i, z_i)\}$, the volume reconstruction process is then given by
\begin{equation}\label{eq:recon}
    V_{n_1,n_2,n_3} = \frac{\Sigma_{i=1}^{N} W(x_i-n_1)W(y_i-n_2)W(z_i-n_3)V_{x_i,y_i,z_i}}{\Sigma_{i=1}^{N} W(x_i-n_1)W(y_i-n_2)W(z_i-n_3)}
\end{equation}
where $(n_1,n_2,n_3)\in \mathbbm{N}^3$ is voxel index, and the weight function is given by
\begin{equation}\label{eq:w}
    W(u)=\mathbbm{1}_{|u|\leq 1}(u)\cdot(1-|u|)
\end{equation}

The indicator function in Eq. \ref{eq:w} suggests that any support data point within one certain cube will only contribute to the eight vertices, or reversely, the value at any query grid point can be calculated based on contributions of support data within adjacent eight cubes. The computation of these contributions thus can be done independently between all these cubes, before gathering all these contributions for estimating the query vertex values, thus $O(N)$\footnote{The interpolation process has an average speed of less than 1 ms over the dataset.}.
 
\subsection{Deformation Estimation using Co-Optimisation}
\label{ssec:Co-Optimisation in 3D Space}

A transformation-predicting network \cite{balakrishnan2019voxelmorph} $f_{def}$ with parameters $\theta_{def}$ takes the interpolated rigid-transformed $\hat{V}$ as input and generate a dense displacement field (DDF) $\phi$:
$\phi = f_{def}(\hat{V};\theta_{def})$.
The DDF then warps the rigid-reconstruction-predicted volume to obtain the final prediction: $\widetilde{V} = \hat{V}\circ \phi$.
As illustrated in Fig.~\ref{pipeline} (a), the deformation estimation process is co-optimised together with rigid reconstruction in Section~\ref{ssec:Reconstruction of stacked 2D frames}. 

The mean squared error, between predicted points coordinates and ground-truth points coordinates, is used to supervise the rigid reconstruction:
$\mathcal{L}_{recon}=\frac{1}{M}\times \frac{1}{N}\times\sum_{m=1}^{M}\sum_{n=1}^{N}\vert \vert P^n_m-\hat{P}^n_m\vert \vert^2_2$,
where $P^n_m$ and $\hat{P^n_m}$ are the points coordinates of ground-truth and prediction respectively, transformed from ground-truth transformation $\textbf{t}^{ref\leftarrow m}$ and predicted transformation $\hat{\textbf{t}}^{ref\leftarrow m}$.
$\textbf{t}^{ref\leftarrow m}$ is calculated using two tool-to-world transformations, $T^{ref\leftarrow m}=(T^{world\leftarrow ref})^{-1} \cdot T^{world\leftarrow m}$, $T$ denoting transformation matrices, converted from parameter vectors. The tool-to-world transformation is obtained from an optical tracker.

The loss function for training the deformation network 
is a typical registration loss \cite{hu2018weakly}, 
consists of bending energy and intensity similarity between the ground-truth-reconstructed volume $V$ and wrapped predicted volume $\widetilde{V}$, to not only encourage a smooth deformation generated by the network, but also provides an opportunity to rectify any erroneous rigid reconstruction estimation:
$\mathcal{L}_{def}(\widetilde{V},V,\phi)=\mathcal{L}_{sim}(\widetilde{V},V)+\mathcal{L}_{smooth}(\phi)$

\subsection{Training Strategy}
\label{ssec:Training Strategy}

In this section, we describe two training strategies that can be used in the proposed pipeline, meta-learning and end-to-end training.

In the meta-learning strategy, network parameters involved in two processes described in Sections ~\ref{ssec:Reconstruction of stacked 2D frames} and ~\ref{ssec:Co-Optimisation in 3D Space} are separately optimised using a training data set $\mathcal{D}_{train}=\{S,\{\textbf{t}_m^{ref\leftarrow m}\}_{m=1}^M\}$ and a validation data set $\mathcal{D}_{val}=\{\widetilde{V},V\}$, respectively:
$\hat{\theta}_{def} = \arg\mathop{\min}_{\theta_{def}} \mathcal{L}_{def}^{val} (\theta_{def}; \hat{\theta}_{recon},\mathcal{D}_{val}),
~\textrm{s.t.}~ \hat{\theta}_{recon} = \arg\mathop{\min}_{\theta_{recon}} \mathcal{L}_{recon}^{train} (\theta_{recon};\mathcal{D}_{train})$.
This bi-level optimisation strategy updates the reconstruction and deformation networks, on separate training and validation sets. This has widely been adopted in previous work, including those for this application \cite{li2023privileged}, to take into account the co-dependency between the optimisation of the two networks and avoid sub-optimum solutions to both optimisations. 

However, we have found that such trivial solutions are unlikely in this application, perhaps due to the highly constrained deformation estimation. Therefore, we propose to use a simple weighted single loss function to co-optimise the two networks in an end-to-end training. The loss functions in Sections ~\ref{ssec:Reconstruction of stacked 2D frames} and ~\ref{ssec:Co-Optimisation in 3D Space} are combined by weighting to train both network parameters $\hat{\theta}_{recon}$ and $\hat{\theta}_{def}$: 
$ \hat{\theta}_{recon},\hat{\theta}_{def} = \arg\mathop{\min}_{\theta_{recon},\theta_{def}} \mathcal{L}^{train}_{ete}(\theta_{recon},\theta_{def};\mathcal{D}_{train})$,
where the end-to-end loss $\mathcal{L}_{ete}^{train}$ used for supervising the co-optimisation process consists of two, with a weight $\alpha$ calculated based on the magnitude of gradient of two parts \cite{lin2021closer}:
$\mathcal{L}_{ete}^{train}=\mathcal{L}^{train}_{recon}+\alpha\times\mathcal{L}^{train}_{def}$.

\subsection{Evaluation Metrics}
\label{ssec:Evaluation Metrics}

Using rigid transformation, e.g. recorded by a spatial tracker, as ground-truth, the weighting between translation and rotation components may be difficult to interpret. In this paper, we design and propose four streamlined evaluation metrics, on pixel and landmark reconstruction error, at local and global levels.

We first define two types of transformation-representing displacement vectors - global displacement vectors and local displacement vectors, where the former represents the displacement between each frame and the reference frame (i.e., the first frame in this work) and the latter denotes the displacement between each frame and the immediately previous frame.

We then define two types of errors, consisting of 1) pixel reconstruction error, where the reconstruction error is the averaged Euclidean distance between ground-truth- and predicted- reconstructed points locations, averaged over all pixels of all but the reference frame in a scan; and 2) landmark reconstruction error, where the reconstruction error is averaged over landmarks in a scan, with the same measurement as pixel reconstruction error.

Thus, the four evaluation metrics used in this study are based on the two types of displacement vectors and the two types of error measurements: 1) global pixel reconstruction error (GPE), reconstruction error on all pixels based on global displacement vectors; 2) global landmark reconstruction error (GLE), reconstruction error on landmarks based on global displacement vectors; 3) local pixel reconstruction error (LPE), reconstruction error on all pixels based on local displacement vectors; and 4) local landmark reconstruction error (LLE), reconstruction error on landmarks based on local displacement vectors. 

The proposed global and local levels of displacement vectors are capable of reflecting the reconstruction error on both frame-level and accumulated error of the algorithm \cite{li2023trackerless}. In addition to the scenario where the performance on the entire scan or adjacent frames is required, as measured by the above metrics, other clinical applications may reconstruct a sequence of US frames using different application-dependent intervals. Nonetheless, these four metrics should still provide an estimate of performance range, for these applications with varying trade-off between accumulated error and reference updating. 

\section{Experiments}
\label{sec:Experiments}

\noindent\textbf{Data Acquisition:} The \textit{in vivo} data 
\footnote{This study was performed in accordance with the ethical standards in the 1964 Declaration of Helsinki and its later amendments or comparable ethical standards. Approval was granted by the Ethics Committee of local institution (UCL Department of Medical Physics and Biomedical Engineering) on $20^{th}$ Jan. 2023 [24055/001].}
used in this paper were acquired from 60 volunteers, using Ultrasonix machine (BK, Europe) with a curvilinear probe (4DC7-3/40), tracked by an NDI Polaris Vicra (Northern Digital Inc., Canada). Other US imaging parameters are empirically configured based on the visual quality of acquired US images. For example, the frequency was set at 6MHz with a depth of 9 cm, and the dynamic range is 83 dB with an overall gain of 48\%. The US frame, with an image size of $640\times480$, was recorded at 20 frames per second (fps), without speckle reduction. The spatial calibration was obtained using a pinhead-based method \cite{hu2017freehand}, and the temporal difference between the optical tracker and imaging was calibrated using the Plus Toolkit~\cite{lasso2014plus}.  
 
Twelve scans were acquired for each subject, from both left and right arms, with the US probe perpendicular of and parallel to the scanning direction, in three different scanning trajectories - straight, c-shape and s-shape, in a distal-to-proximal direction, resulting in 720 scans in total.
The average number of frames per scan is 500, equivalent to $200-300$ mm.  
The data set was split into train, validation and test sets by a ratio of 3:1:1 on subject level, where scans from the same subject cannot be in different sets.

\noindent\textbf{Network Development and Implementation:}
EfficientNet (b1) \cite{tan2019efficientnet} is adopted as the backbone of the reconstruction network. A fully connected layer is added at the end to output $(M-1)\times6$ rigid transformation parameters,
with US sequence containing $M$ frames as input. After calculating the points locations in real-world space for pixels in the input US sequence and interpolating into an US volume, the rigid-transformed volume is fed into an adapted VoxelMorph \cite{balakrishnan2019voxelmorph} network, with an input channel of 1.

The input sequence length is set to 100 for reported benchmark performance~\cite{10288201}.
The reconstructed volume has a resolution of 1 mm$\times$1 mm$\times$1 mm with various sizes.
Other hyper-parameters are less sensitive to reconstruction and selected based on validation performance, including an Adam optimizer with a learning rate of $10^{-4}$ and $\alpha=10^{3}$, on a single NVIDIA Quadro GV100 GPU.

For a fair comparison, we adapted two start-of-the-art (SOTA) methods, \cite{prevost20183d} and \cite{li2023trackerless}, with the same hyper-parameters.
We also compared with a model with meta-training strategy, adapted from \cite{liu2018darts}. The same data split was used with a ratio of 2:2:1, where the train and validation sets for rigid transformation and nonrigid deformation network training, respectively.
All models were trained for at least 10,000 epochs until convergence, for up to 5 days.

\section{Results}
\label{sec:Results}
The performance improvement from our proposed method is summarised in Table~\ref{metrics}, as well as results from an ablation study with only rigid transformation prediction (baseline) or meta-trained (\textit{Meta}) models.
As an surrogate of clinically useful landmarks, four corner points in an image were used.
\textit{Recon} and \textit{Def} models come from our proposed approach, using the co-optimised rigid transformation and nonrigid transformation, respectively.

The reconstruction performance of the \textit{Recon} and \textit{Def} models both show performance improvement in global reconstruction metrics (GPE and GLE), compared with the baseline method ($p$-value $=0.001$ and $0.003$ for \textit{Def} model, based on paired t-test at a significance level at $\alpha=0.05$). The performance of \textit{Meta} model is poorer than those from both the end-to-end model and the baseline, for metrics, demonstrating the effectiveness of proposed end-to-end  
\begin{table}
\begin{center}
\caption{Reconstruction performance using proposed four evaluation metrics, among baseline, SOTA and our methods.}\label{metrics}%
\begin{tabular*}{\textwidth}{@{\extracolsep{\fill}}ccccccc@{\extracolsep{\fill}}}
\hline
Models & GPE (mm) & GLE (mm) & LPE (mm)& LLE(mm)  \\
\hline

Baseline & $18.48\pm10.30$ &    $19.70\pm10.42$ &    $\boldsymbol{0.41\pm0.17}$ &  $\boldsymbol{0.44\pm0.18}$ \\
Recon\textsubscript{meta} & $21.29\pm11.01$    &  $22.82\pm11.15$  &  $3.59\pm1.44$ &  $3.71\pm1.45$ \\
Def\textsubscript{meta} & $20.89\pm10.52$ &  $22.80\pm11.13$ & --  & --\\
Recon (ours) & $16.69\pm7.79$ & $18.15\pm7.91$ & $3.07\pm0.99$  & $3.38\pm1.00$  \\
Def (ours) & $\boldsymbol{16.51\pm7.76}$ & $\boldsymbol{17.91\pm7.85}$  & --  & --\\

\hline

\cite{prevost20183d}\textsubscript{cf} & $18.33\pm7.77$ &  $20.19\pm8.04$ &  $\boldsymbol{0.23\pm0.07}$ &  $\boldsymbol{0.25\pm0.08}$  \\
\cite{li2023trackerless}\textsubscript{cf} & $17.32\pm8.12$ &  $18.64\pm8.53$ &  $0.23\pm0.08$ &  $0.25\pm0.09$ \\
Baseline\textsubscript{cf} & $16.44\pm7.83$ &   $17.75\pm8.08$ &    $0.39\pm0.16$ &  $0.44\pm0.18$\\ 
Recon\textsubscript{meta_cf}& $18.46\pm8.77$  &  $19.93\pm8.95$ &  $3.29\pm1.37$ &  $3.57\pm1.48$ &  &  \\
Def\textsubscript{meta_cf} &  $18.21\pm8.37$  &  $19.94\pm8.90$ & -- & -- \\
Recon\textsubscript{cf} (ours) &  $15.26\pm7.17$ &    $16.57\pm7.43$ &    $2.71\pm0.89$ &  $3.25\pm0.97$ \\ 
Def\textsubscript{cf} (ours) &  $\boldsymbol{15.13\pm7.12}$ & $\boldsymbol{16.37\pm7.35}$ & -- & --\\

\hline
\end{tabular*}
\end{center}
\end{table}
strategy. Most interestingly, although the global reconstruction metrics show significant improvement, the local metrics has a reduced performance. This may reflect a property of the added regularised deformation estimation, in which the local perturbation (albeit may yield larger variance) is constrained to have a smaller bias, thus reducing the long-term (or long distance in this case) expected error. 

\begin{figure}[t]
 \centering
\includegraphics[width=\textwidth]{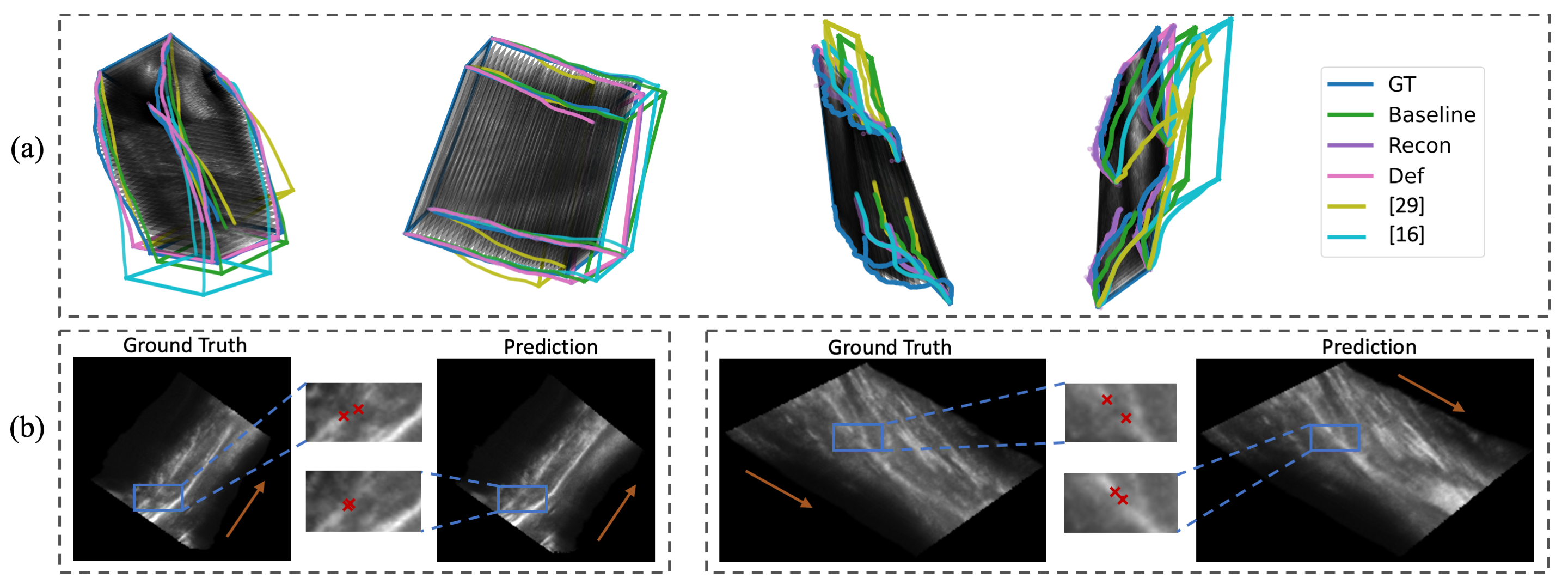}
\caption{(a): Reconstructed US scan volumes with ground-truth, proposed and SOTA methods, illustrated with perpendicular c-shape, perpendicular straight, parallel s-shape and parallel c-shape scans, from left to right. (b) Illustration of landmarks in ground-truth- and prediction- reconstructed US scan volumes.} \label{recon}
\end{figure}

The number of frames in an input US sequence is set at $100$ in \cite{li2023trackerless}, with transformation from $20^{th}$ to $21^{th}$ frames as an example. As the method in \cite{li2023trackerless} cannot predict the probe trajectory for all frames in a scan. For fair comparison, we subsample the frames and use the same reference frame for the other methods in Table~\ref{metrics}, so that all methods predict the transformation for the same subset of frames, i.e. the “common frames”, denoted as \textsubscript{cf}.
When comparing on common frames between proposed and baseline models, the improvement was also observed with $p$-value $=0.004$ and $0.003$ in GPE and GLE, using \textit{Def} model, also illustrated in Fig.~\ref{recon} (a).

The proposed co-optimisation approach has potentials to rectify and improve rigid reconstruction with ground-truth, by compensating nonrigid deformation.
Fig.~\ref{recon} (b) shows two example slices on the same location, from ground-truth-reconstructed volumes and DDF-predicted volumes, arrow indicating the scanning direction. 
The landmarks (red crosses) in each slice represent the same anatomical structure, and thus should be at the same 3D location when reconstructed. It can be seen that the anatomical structure is broken in the ground-truth-reconstructed volume, perhaps due to movement or tissue motion. The distance between these landmarks decreased in the predicted volume, from an average distance of $4.65$ mm to $1.48$ mm.

\section{Conclusion and Discussion}

This work introduced deformation estimation into rigid reconstruction of freehand US. The experimental results, evaluated on a large data set, shows the efficacy of the proposed algorithm. Examples of compensating nonrigid deformation are also discussed, to open up new avenue for improving this longstanding challenge in ultrasound image computing. 

\begin{credits}
\subsubsection{\ackname} This work was supported by the EPSRC [EP/T029404/1], a Royal Academy of Engineering / Medtronic Research Chair [RCSRF1819\textbackslash7\textbackslash734] \makebox[\textwidth][s]{(TV), Wellcome/EPSRC  Centre  for Interventional  and  Surgical  Sciences} \newline[203145Z/16/Z], and the International Alliance for Cancer Early Detection, an alliance between Cancer Research UK [C28070/A30912; C73666/A31378], Canary Center at Stanford University, the University of Cambridge, OHSU Knight Cancer Institute, University College London and the University of Manchester. TV is co-founder and shareholder of Hypervision Surgical. Qi Li was supported by the University College London Overseas and Graduate Research Scholarships. For the purpose of open access, the authors have applied a CC BY public copyright licence to any Author Accepted Manuscript version arising from this submission.

\end{credits}

\bibliographystyle{splncs04}
\bibliography{Paper}
\end{document}